\documentclass[submission, Phys]{SciPost}
\usepackage[english]{babel}
\usepackage{amsthm}
\usepackage{amsmath}
\usepackage{amssymb}
\usepackage{xcolor}
\usepackage{yhmath}
\usepackage[mathscr]{euscript}
\usepackage{titlesec}
\theoremstyle{definition}

\theoremstyle{remark}

\usepackage{comment}

\newcommand\blfootnote[1]{%
  \begingroup
  \renewcommand\thefootnote{}\footnote{#1}%
  \addtocounter{footnote}{-1}%
  \endgroup
}
\begin{document}

\begin{center}{\Large \textbf{
Path Integral approach to Black-hole Evaporation and Accretion
}}\end{center}

\begin{center}
I. Arraut\textsuperscript{1$\dagger$}, A. K. Mehta\textsuperscript{2}
\end{center}

\begin{center}
{\bf 1}Institute of Data Engineering and Science (IDEAS) \&
 Institute of Science and Environment (ISE),
 University of Saint Joseph, Macao
\\
*ivan.arraut@usj.edu.mo\\
{\bf 2}Department of Physics, Kyunghee University, Seoul, Republic of Korea
\\
*abhishek.mehta@khu.ac.kr
\end{center}

\begin{center}
\today
\end{center}

\section*{Abstract}
{\bf  In this paper, we investigate evaporation and accretion of uncharged, non-rotating, spherically symmetric black-holes from the path integral perspective. We show that the effective actions derived using the path integral techniques incorporate both accreting and evaporating configurations, thus presenting a novel methodology to study black-hole accretion and evaporation in $4D$ on the same footing. We specifically focus on evaporating configurations which deviate from the standard thermodynamic modes of black-hole evaporation. 


}

\vspace{10pt}
\noindent\rule{\textwidth}{1pt}
\tableofcontents\thispagestyle{fancy}
\noindent\rule{\textwidth}{1pt}
\vspace{10pt}
\blfootnote{${}^{\dagger}$Corresponding author}
\section{Introduction}

Black-holes are compact objects generated at the end of a stellar life-cycle when a large amount of mass is concentrated behind an event horizon \cite{1}. The existence of an event horizon endows the black-hole with many strange properties which become apparent in the semi-classical regime of gravity. One of the most celebrated being that black-holes emit radiation \cite{Hawking:1975vcx}. As a consequence, black-holes undergo decay over time, known as black-hole evaporation, which is also related to a host of other consequences that is collectively addressed as black-hole thermodynamics. One can compute the rate of black-hole evaporation by solving the following field equation for the back-reaction\cite{Ford:2005qz}
\begin{align}
    G_{\mu\nu} = 8\pi \langle T_{\mu\nu}\rangle \label{br}
\end{align}
where in the above the expectation value is computed over the matter fields only. If one works with the following metric ansatz
\begin{align}
    ds^2 = -\left(1-\frac{2M(r, t)}{r}\right)dt^2+\frac{dr^2}{1-\frac{2M(r, t)}{r}}+r^2d\Omega^2_{S^2}.\label{metans}
\end{align}
Then the $rt$-component of the above semi-classical field equation gives
\begin{align}
&\frac{\partial M(r, t)} {\partial t} = 4\pi r^2\left(1-\frac{2M(r, t)}{r}\right)\langle T_{rt}\rangle\notag\\
&\implies \lim_{r\to\infty}\frac{\partial M(r, t)} {\partial t} = 4\pi\lim_{r\to\infty}\int  d\Omega_{S^2}~\sqrt{g}\langle T_{rt}\rangle.\label{mlrate}
\end{align}
Notice that this is exactly the formula that is used in obtaining the rate of mass loss of a black-hole due to Hawking radiation\cite{Unruh:1974bw, dewitt1975quantum}. At first glance, it is not obvious that the formula will yield an evaporating configuration for any matter field, let alone an evaporation that is thermodynamically-mediated. Indeed, the above formula presents us with the classic thermodynamically-mediated Hawking radiation only for conformally-coupled and massless scalar fields\cite{Unruh:1974bw, Hawking:1975vcx, Birrell:1978ng}. For other classes of scalar field theories that have mass $m$, self-interactions with coupling parameter $\lambda$ and/or have nonminimal coupling with gravity with coupling parameter $\xi$, Eq. \ref{mlrate} will show explicit dependence on these parameters which may not result in a thermodynamically-mediated evaporating configuration for all arbitrary values of $m$, $\lambda$ and $\xi$. This is precisely what was observed for the case of scalar fields with quartic interactions in $2D$ \cite{Leahy:1983vb} where the thermal and non-thermal nature of black-hole evaporation was dependent on the thermal and non-thermal nature of the ingoing and outgoing vacuum states. Therefore, non-thermodynamical black-hole evaporation is to be expected when generic interactions are involved. Moreover, most computations in $2D$ gravity that show the thermodynamic nature of black-holes are heavily dependent on the choice of quantization procedure, the choice of incoming and outgoing vacua, regularization procedure and the definition of particle states in non-flat spacetimes\cite{unruh1976notes, davies1976energy, hooft1985quantum, Diatlyk:2020nxa} that are made on the basis of physical reasonability. It is possible that non-thermodynamic modes of black-hole evaporation are getting ignored in these assumptions. In such a scenario, the best way to see all possible modes of black-hole evaporation is through path integral-based effective field theory techniques. In fact, Eq. \ref{br} itself is implicitly an equation of motion of an effective gravitational field theory. To see this, notice that
\begin{align}
    T_{\mu\nu} = \frac{1}{\sqrt{g}}\frac{\delta S_{matter}[g, \varphi]}{\delta g^{\mu\nu}}.
\end{align}
where $\varphi$ are matter fields. The above can be used to rewrite Eq. (\ref{br}) as follows
\begin{align}
    G_{\mu\nu} = \langle T_{\mu\nu}\rangle &= \frac{1}{\sqrt{g}}\left\langle \frac{\delta S_{matter}[g, \varphi]}{\delta g^{\mu\nu}}\right\rangle \notag\\&= \frac{1}{Z[g]}\frac{1}{\sqrt{g}}\int  [D\mathcal{\varphi}]\frac{\delta S_{matter}[g, \varphi]}{\delta g^{\mu\nu}}e^{-iS_{matter}[g, \varphi]} \notag\\&= \frac{i}{Z[g]}\frac{1}{\sqrt{g}}\frac{\delta}{\delta g^{\mu\nu}}\int  [D\mathcal{\varphi}]e^{-iS_{matter}[g, \varphi]}  , \label{PI}
\end{align}
where
\begin{align}
    Z[g] \equiv \int  [D\mathcal{\varphi}]e^{-iS_{matter}[g, \varphi]}
\end{align}
 For simplicity, let's assume that $S_{matter}$ is given by the scalar field theory in curved background
\begin{align}
    S_{matter}[g, \varphi] = \int d^4x \frac{\sqrt{g}}{2}(\nabla_{\mu}\varphi\nabla^{\mu}\varphi+m^2\varphi^2
    ).
\end{align}
Then in Eq. \ref{PI}, the last equality is just the path integral of the matter fields and is given by 
\begin{align}
 \int [\mathcal{D}\varphi] e^{-iS_{matter}[g, \varphi]} = {\det} ^{-1/2}(-\Box+m^2) = \exp\left(-\frac{1}{2}\int \operatorname{Tr}\ln(-\Box+m^2)\right).
\end{align}
Using the heat kernel method\cite{Vassilevich:2003xt,Decanini:2005gt} it can be shown that
\begin{align}
    &\operatorname{Tr}\ln(-\Box+m^2) \notag\\&= \sqrt{g}(1 + R\cdot R + R\cdot R\cdots R + \nabla R \cdot\nabla R + R\nabla\cdot\nabla R\cdots\nabla\cdots R\nabla\cdots\nabla R + \cdots \equiv S_{eff}[R, R_{\mu\nu}, R_{\mu\nu\rho\sigma}]\label{effG}
\end{align}
which is just Einstein-Hilbert action modified via higher curvature terms. Using the above in Eq. (\ref{PI}) then leads to
\begin{align}
    \langle T_{\mu\nu}\rangle = \frac{1}{\sqrt{g}}\frac{\delta }{\delta g^{\mu\nu}}S_{eff}[R, R_{\mu\nu}, R_{\mu\nu\rho\sigma}]
\end{align}
Therefore, equivalently, Eq. \ref{br} is the equation of motion for EH gravity modified by higher curvature terms. After making use of the metric ansatz in Eq. \ref{metans} in the effective action in Eq. \ref{effG}, one can obtain an effective action for the mass parameter $M(r,t)$. The equation of motion to this effective action in principle should give us all modes of black-hole evaporation. 
\section{Effective action for the black-hole mass}
\markboth{EFFECTIVE ACTION FOR\\ THE BLACK-HOLE MASS}
~
In practice, however, it turns out that computing $S_{eff}[R, R_{\mu\nu}, R_{\mu\nu\rho\sigma}]$ using the heat kernel method is quite tedious and cannot be realistically performed beyond a few orders\cite{Decanini:2005gt} and the ansatz in Eq. \ref{metans} is also a bit too complicated for this purpose. Hence, we make the following changes to our computational strategy. We write the action of the scalar field $\varphi$ in the background of the Vaidya metric and then integrate out the scalar field to obtain the effective action for $M(u)$. Both methods are equivalent and give the same effective action for the mass parameter. Then, we shall study the equation of motion of this effective action.\\\\ Therefore, let us now look at the Klein-Gordon action in the background of the Vaidya metric. The Vaidya metric is given by
\begin{align}
    &g^{\mu\nu} = 
    \begin{pmatrix}
        0 & -1  & 0 & 0\\
        -1 & \left(1-\frac{2M(u)}{r}\right) & 0 & 0\\
        0 & 0  & r^{-2} & 0\\
        0 & 0 & 0 & r^{-2}\sin^{-2}\theta
    \end{pmatrix}  = \eta^{\mu\nu}_{NC} -\frac{2M(u)}{r}\delta^{\mu}_r\delta^{\nu}_r,\\
    &g_{\mu\nu} = (\eta_{NC})_{\mu\nu}+\frac{2M(u)}{r}\delta^u_{\mu}\delta^u_{\nu} \quad \sqrt{g} = \sqrt{\eta_{NC}},
\end{align}
where $\eta^{\mu\nu}_{NC}$ refers to the Minkowski metric in the null coordinates given by
\begin{align}
    ds^2_{NC} = (\eta_{NC})_{\mu\nu}dx^{\mu}dx^{\nu} = -2dudr-du^2 + r^2d\Omega^2_{S^2}.
\end{align}
Now, the KG action may be written as
\begin{align}
    S[M, \varphi] &= \int_{\mathcal{M}} d^4x \sqrt{\eta_{NC}}[\partial^{\mu}\varphi\partial_{\mu}\varphi+m^2\varphi^2-\frac{2M(u)}{r}(\partial_r\varphi)^2]\label{KGAct} 
\end{align}
We now expand the scalar field in the following spherically symmetric basis 
\begin{align}
     r^2\vec{\nabla}^2\psi_{l,m}(k)  = -k^2 r^2 \psi_{l,m}(k)\label{ssymmbasis}
\end{align}
However, due to the event horizon at $r = 2M(u)$ being a one-way permeable membrane, the only consistent set of spherically symmetric basis functions in presence of such a boundary is given by\footnote{See Appendix \ref{movBound}}
\begin{align}
 r^2\vec{\nabla}^2\psi_{l,m}  = 0.  
\end{align}
which is just equivalent to setting $k=0$ in Eq. \ref{ssymmbasis}. Therefore, one can also physically interpret the above as a result of an infinite redshift at the event horizon. Assuming azimuthal symmetry for simplicity, we will work with $m=0$ so that we have
\begin{align}
    &\psi^{in}_{l}(u, r, \theta) = \sqrt{\frac{2l+1}{2M(u)}}
    \left(\frac{r}{2M(u)}\right)^{l}\widetilde{P}_l(\cos\theta) \quad r < 2M(u),\notag\\
    &\psi^{out}_{l}(u, r, \theta) =  \sqrt{\frac{2l+1}{2M(u)}} \left(\frac{r}{2M(u)}\right)^{-l-1}\widetilde{P}_l(\cos\theta)\quad r > 2M(u),\label{sphmodefn}
\end{align}
where $\widetilde{P}_l(z)$ is the normalized Legendre polynomial such that
\begin{align}
    &\psi^{in}_{l}(u, 2M(u), \theta) = \psi^{out}_{l}(u, 2M(u), \theta),\label{modeCons}\\
    &\int^{2M(u)}_0 dr d\Omega_{S^2} |\psi^{in}_{l}(u, r, \theta)|^2 = \int^{\infty}_{2M(u)} dr d\Omega_{S^2} |\psi^{out}_{l}(u, r, \theta)|^2 = 1.
\end{align}
Hence, we may write
\begin{align}
    \varphi(u, r, \theta) =  \sum_{l} \left[a_{l}(u)\psi^{in}_{l}(u, r, \theta)+b_{l}(u)\psi^{out}_{l}(u, r, \theta)\right]\label{RIMo}
\end{align}
The above splits the action as
\begin{align}
    S = S_{in} + S_{out}.
\end{align}
which when evaluated leads to
\begin{align}
    S_{in} &= \sum_{l}\int du  ~a_{l}(u)^2(2l+1)\left[2l\dot{M} + \frac{4m^2}{2l+3}M^2-\frac{l}{2}\right],\\
    S_{out} &= \sum_{l}\int du  ~b_{l}(u)^2(2l+1)\left[2(l+1)\dot{M} + \frac{4m^2}{2l-1}M^2-\frac{l+1}{2}\right].
\end{align}
Now, we compute the following path integral for the action in Eq. \ref{KGAct} following\cite{Lee:2021iut}
\small
\begin{align}
    Z[M]  &= \int [\mathcal{D}\varphi] e^{iS[M, \varphi]}\notag\\&= \int [\mathcal{D}a_{l}][\mathcal{D}b_{l}] e^{i(S_{in}+S_{out})} = \prod_{u\in \mathbb{R}} \prod_{l\in \mathbb{Z}}\left(\frac{2l+1}{2l+3}\right)^{-1/2}(4\dot{M}-1)^{-1/2}\left[l+z_{+}(\dot{M}, M)\right]^{-1/2}\left[l+z_{-}(\dot{M}, M)\right]^{-1/2} \notag\\
    &=\prod_{u\in\mathbb{R}}\frac{1}{\sqrt{\sin[\pi z_{+}(\dot{M}, M)]\sin[\pi z_{-}(\dot{M}, M)]}} =\prod_{u\in\mathbb{R}}\frac{1}{\sqrt{\cos[2\pi f(\dot{M}, M)]}} = e^{iS_{eff}[M]},\\
    &z_{\pm}(\dot{M}, M) = \frac{3}{4}\pm\frac{1}{4}\sqrt{\frac{36\dot{M}-64m^2M^2-9}{4\dot{M}-1}} \equiv  \frac{3}{4}\pm f(\dot{M}, M),
\end{align}
\normalsize
where in the second equality we have made a change of the path integral measure using the mode expansion in Eq. \ref{RIMo}. The effective action for $M$ can be read off from the above and is given by
\begin{align}
    S_{eff}[M] =-\frac{1}{2} \int du \ln\cos[2\pi f(\dot{M}, M)].
\end{align}
The equation of motion of the above is given by
\begin{align}
    &2\pi\sec^2(2\pi f)\frac{df}{du}\frac{\partial f}{\partial \dot{M}} + \tan(2\pi f)\left[\frac{d}{du}\left(\frac{\partial f}{\partial \dot{M}}\right)-\frac{\partial f}{\partial M}\right] = 0\label{EOMNoTh0}\\
    & \frac{df}{du} \equiv -\frac{f \ddot{M}}{2(4\dot{M}-1)} + \frac{9\ddot{M}-32m^2M\dot{M}}{8(4\dot{M}-1)f}\label{EOMNoTh}.
\end{align}
Also, notice due to the Vaidya metric, Eq. (\ref{PI}) becomes
\begin{align}
     &\langle T_{\mu\nu}\rangle = \frac{1}{\sqrt{g}}\frac{\delta }{\delta g^{\mu\nu}}S_{eff}[R, R_{\mu\nu}, R_{\mu\nu\rho\sigma}] = -\frac{r}{\sqrt{\eta_{NC}}\delta^{\mu}_r\delta^{\nu}_r}\frac{\delta S_{eff}[M]}{\delta M}\\
     &\implies \langle T_{rr}\rangle = -\frac{r}{\sqrt{\eta_{NC}}}\frac{\delta S_{eff}[M]}{\delta M}
\end{align}
where in the above we have also utilized the fact that the curvature tensors are now only functionals of $M(u)$. Due to the Vaidya metric, we also have $\langle T_{rr}\rangle = G_{rr} = 0$, therefore, the above becomes
\begin{align}
    \frac{\delta S_{eff}[M]}{\delta M} = 0
\end{align}
Hence, due to the Vaidya metric, the semiclassical, back-reaction equation in Eq. (\ref{mlrate}) completely equivalent to the equation of motion of the path-integral derived effective action $S_{eff}[M]$ demonstrating the equivalence between the two approaches. Now, we will try to guess a solution for the above. In Eq. \ref{EOMNoTh}, we can have the second term vanish by
\begin{align}
    &\ddot{M} = \frac{32}{9}m^2M\dot{M}\notag\\
    &\implies \dot{M} = c + \frac{16 m^2 M^2}{9} \label{NTErate}
\end{align}
where $c$ is an integration constant. Notice that we can rewrite the above as
\begin{align}
    f = \frac{3}{2}\sqrt{\frac{c-1/4}{4\dot{M}-1}}
\end{align}
By setting $c = 1/4$, we get $f = 0$ which also leads to the vanishing of the first term in Eq. \ref{EOMNoTh} as well. Since, $\frac{df}{du} = 0$ and $f = 0$, Eq. \ref{EOMNoTh0} is also satisfied. Hence, we have
\begin{align}
    \dot{M} &= \frac{1}{4} + \frac{16 m^2 M^2}{9} \\&\overset{\text{large}~M}{\approx} \frac{16 m^2 M^2}{9} \label{largeM}
\end{align}
Notice that Eq. \ref{NTErate} for $m^2 > 0$ describes a cascade not an evaporation. In fact, for large $M$ this is precisely the mass dependence observed in the Bondi Accretion formula \cite{bondi1952spherically} for spherically symmetric accretion which is given by
\begin{align}
    \dot{M} = \frac{\pi \rho G^2 M^2}{c^3_s}
\end{align}
where $\rho$ is the ambient density and $c_s$ is the speed of sound in that medium. By restoring the gravitational constant $G$ in Eq. \ref{largeM}, we can make the following identification
\begin{align}
    m^2 \to \frac{9G\pi \rho}{16c_s^3}
\end{align}
This implies that the medium of density $\rho$ accreting into the black-hole can be effectively understood as a scalar field of mass $\sqrt{\frac{9G\pi \rho}{16c_s^3}}$ in a black-hole background.
\subsection{Existence of remnant configurations}

When $m^2 < 0$, $M_r = \frac{3}{8|m|}$ is a critical point and it satisfies the stability criteria $d\dot{M}/d M < 0$. The solution to the above is given by
\begin{align}
    &M(u) = M_r\coth\left[\frac{u}{4M_r}+\coth^{-1}\left(\frac{M_i}{M_r}\right)\right] \quad M_i > M_r,\\
    &M(u) = M_r\tanh\left[\frac{u}{4M_r}+\tanh^{-1}\left(\frac{M_i}{M_r}\right)\right] \quad M_i < M_r,\\ 
    &M_r=\frac{3}{8|m|},
\end{align}
where $M_{i}$ is the initial mass of the black-hole. When $M > M_r$ we have a decaying black-hole and a cascading one when $M < M_r$. A cascading solution implies that the black-hole is consuming matter instead of emitting it as radiation. However, notice that when $u\to\infty$ the black-hole mass saturates to $M_r$. Therefore, the stable point of the above ODE has the interpretation of a black-hole remnant.
\subsection{Without azimuthal symmetry}
Here, we repeat the path integral computation without assuming azimuthal symmetry and work with general spherical harmonics
\begin{align}
r^2\vec{\nabla}^2\psi_{l,m }  = 0,  
\end{align}
so that we have
\begin{align}
    &\psi^{in}_{l,m}(u, r, \theta,\phi) = \sqrt{\frac{2l+1}{2M(u)}}\left(\frac{r}{2M(u)}\right)^{l}Y^{m}_{l}(\theta,\phi) \quad r < 2M(u),\\
    &\psi^{out}_{l,m}(u, r, \theta,\phi) =   \sqrt{\frac{2l+1}{2M(u)}}\left(\frac{r}{2M(u)}\right)^{-l-1}Y^{m}_{l}(\theta,\phi) \quad r > 2M(u).
\end{align}
Hence, we may write
\begin{align}
    &\varphi^{out}(u, r, \theta,\phi) = b_{0,0}(u)\psi^{out}_{0,0}(u,r,\theta,\phi) + \sum_{\substack{l\neq 0\\m < |l|}}[b_{l, m}(u)\psi^{out}_{l,m}(u,r,\theta,\phi)+\bar{b}_{l,m}(u)\bar{\psi}^{out}_{l,m}(u,r,\theta,\phi)],\\
    &\varphi^{in}(u, r, \theta,\phi) = a_{0,0}(u)\psi^{in}_{0,0}(u,r,\theta,\phi) +\sum_{\substack{l\neq 0\\m < |l|}}[a_{l, m}(u)\psi^{in}_{l,m}(u,r,\theta,\phi)+\bar{a}_{l,m}(u)\bar{\psi}^{in}_{l,m}(u,r,\theta,\phi)],
\end{align}
such that
\begin{align}
    \psi^{in}_{l}(u, 2M(u), \theta, \phi) = \psi^{out}_{l}(u, 2M(u), \theta, \phi).
\end{align}
Using it in the above action, we get
\begin{align}
    S_{in} &= 2\sum_{l,m}\int du  ~|a_{l,m}(u)|^2(2l+1)\left[2l\dot{M} + \frac{4m^2}{2l+3}M^2-\frac{l}{2}\right],\\
    S_{out} &= 2\sum_{l,m}\int du  ~|b_{l,m}(u)|^2(2l+1)\left[2(l+1)\dot{M} + \frac{4m^2}{2l-1}M^2-\frac{l+1}{2}\right].
\end{align}
This time in the path integral, we have
\begin{align}
    Z[M] &= \int [\mathcal{D}a_{l,m}][\mathcal{D}b_{l,m}] e^{iS} \notag\\&= \prod_{u\in \mathbb{R}}\left[\frac{2m^2M^2}{3}(1-4\dot{M})^{2/3}(4\dot{M}-8m^2M^2-1)\right]^{1/2} \notag\\&~~~~~~~~\prod_{l\in \mathbb{Z}}\left(\frac{2l+1}{2l+3}\right)^{-|2l+1|}\left[l+z_{+}(\dot{M}, M)\right]^{-|2l+1|}\left[l+z_{-}(\dot{M}, M)\right]^{-|2l+1|} \notag\\
    &=\prod_{u\in\mathbb{R}}\frac{\left[\frac{2m^2M^2}{3}(1-4\dot{M})^{2/3}(4\dot{M}-8m^2M^2-1)\right]^{1/2} }{\cos[2\pi f(\dot{M}, M)]}\frac{e^{4\left[\zeta(1)-1-\gamma\right](z^2_{+}+z^{2}_{-}-1/2)-2+8\zeta'(-1)}}{G(1+z_{+})^2G(2-z_{+})^2G(1+z_{-})^2G(2-z_{-})^2}\label{PInA}
\end{align}

where we have made use of 
\begin{align}
    \prod_{n=1}^{\infty}\left(1+\frac{z}{n}\right)^n e^{-z+z^2 /(2 n)}=\frac{G(z+1)}{(2 \pi)^{z / 2}} e^{\left[z(z+1)+\gamma z^2\right] / 2},
\end{align}
in the above\footnote{$G(z)$ is the Barnes $G$-Function.}. We now look at
\begin{align}
    \ln Z[M] &= \underbrace{4[\zeta(1)-1-\gamma](z^2_{+}+z^{2}_{-}-1/2)\vphantom{\ln\left(\frac{2m^2M^2}{3}\right)}}_{\Gamma_{div}}+\underbrace{\ln\left(\frac{2m^2M^2}{3}\right)}_{\Gamma_{log}}+\ln\left(\frac{\left[(1-4\dot{M})^{2/3}(4\dot{M}-8m^2M^2-1)\right]^{1/2} }{\cos[2\pi f(\dot{M}, M)]}\right)\notag\\&-2\ln[G(1+z_{+})G(2-z_{+})G(1+z_{-})G(2-z_{-})]-2+8\zeta'(-1) 
\end{align}
Notice that the presence of $\zeta(1) \equiv \epsilon^{-1}$ means that the path integral above is divergent. These are precisely the UV divergences that often appear in effective actions \cite{Vassilevich:2003xt}. These UV divergences can be removed by performing local field redefinitions of $M$ of the form \cite{Solodukhin_2016}
\begin{align}
    M(u, \epsilon) = M_0(u) + \sum_{k}\alpha_k(\epsilon)\mu_k(u)
\end{align}
where $\alpha_k$ is some appropriate function of $\epsilon$, $\mu_k(u)$ is appropriately chosen to cancel the UV divergences . However, we are only interested in the leading behaviour of $M_0$ which is simply the solution to the equation of motion $\Gamma_{div}$ given by
\begin{align}
    \Gamma_{div} \propto  \int du ~f(\dot{M}, M)^2 \label{EAnA}
\end{align}
The equation of motion for the above is 
\begin{align}
4\frac{d}{du}\left(\frac{M}{4\dot{M}-1}\right)+1 = 0.
\end{align}
The solution to the above is
\begin{align}
    M_0(u) =\frac{u+4M_i}{8}+\frac{2M^2_i}{u+4M_i},
\end{align}
which is again a cascading solution. In other words, this is a nontrivial accretion deviating from the usual Bondi Accretion formula. Such accretion is possible when matter accreting into the black-hole has nonzero angular momentum. This is not surprising as the effective action in Eq. \ref{EAnA} is the result of integration over all angular momentum modes in the path integral in Eq. \ref{PInA}. 
\section{Restoring the thermodynamics}
In this section, we will reproduce the standard black-hole evaporation rate as computed by Hawking. For that, we consider the following scalar action
\begin{align}
    S[\varphi] = &\int_{\mathcal{M}} d^4x \sqrt{\eta_{NC}}[\partial^{\mu}\varphi\partial_{\mu}\varphi-\frac{2M(u)}{r}(\partial_r\varphi)^2] +\int_{\mathcal{M}}d^4x\sqrt{g}~\xi(R^2-4R_{\mu\nu}R^{\mu\nu}+R_{\mu\nu\rho\sigma}R^{\mu\nu\rho\sigma})\varphi^2\label{SGBA}
\end{align}
Before we reproduce Hawking's result, we will go through the standard path-integral treatment for completeness. Using Eq. \ref{RIMo} in the above, we will obtain
\begin{align}
    S_{in} &= 2\sum_{l}\int du  ~a_{l}(u)^2(2l+1)\left[2l\dot{M} + \frac{3\xi}{2l-3}M^{-2}-\frac{l}{2}\right],\\
    S_{out} &= 2\sum_{l}\int du  ~b_{l}(u)^2(2l+1)\left[2(l+1)\dot{M} + \frac{3\xi}{2l+5}M^{-2}-\frac{l+1}{2}\right].
\end{align}
Again in the above, we have assumed azimuthal symmetry. Now, we compute the following path integral
\begin{align}
    Z[M] &= \int [\mathcal{D}a_{l}][\mathcal{D}b_{l}] e^{iS} \notag\\&= \prod_{u\in \mathbb{R}} \prod_{l\in \mathbb{Z}}\left(\frac{2l+1}{2l-3}\right)^{-1}(4\dot{M}-1)^{-1/2}\left[l+w_{+}(\dot{M}, M)\right]^{-1/2}\left[l+w_{-}(\dot{M}, M)\right]^{-1/2} \notag\\
    &=\prod_{u\in\mathbb{R}}\frac{1}{\sqrt{\sin[\pi w_{+}(\dot{M}, M)]\sin[\pi w_{-}(\dot{M}, M)]}} =\prod_{u\in\mathbb{R}}\frac{1}{\sqrt{\cos[2\pi g(\dot{M}, M)]}} = e^{iS_{eff}[M]},\\
    &w_{\pm}(\dot{M}, M) = \frac{3}{4}\pm\frac{1}{4}\sqrt{\frac{36\dot{M}-48\xi M^{-2}-9}{4\dot{M}-1}} \equiv  \frac{3}{4}\pm g(\dot{M}, M),
\end{align}
where the effective action for $M$ is given by
\begin{align}
    S_{eff}[M] = \frac{1}{2}\int du\ln\cos[2\pi g(\dot{M}, M)].
\end{align}
From Eq. \ref{EOMNoTh}, it is easy to see that $g = 0$ is a solution to the above which leads to
\begin{align}
    \dot{M} = \frac{1}{4}+\frac{4\xi}{3}M^{-2}.
\end{align}
Notice that for $\xi < 0$ the above does correspond to a thermodynamic evaporation for black-holes of sufficiently small masses. This is an indication that the Scalar-Gauss-Bonnet (SGB) interaction term introduced in Eq. (\ref{SGBA}) is precisely what we require to reproduce standard Hawking's result. When $\xi < 0$, we have a critical point given by $M_c = 4\sqrt{\frac{|\xi|}{3}}$. But notice that
\begin{align}
    \frac{d\dot{M}}{dM}\bigg|_{M = M_c} = \frac{1}{2M_c} > 0,
\end{align}
which implies that $M=M_c$ is an unstable critical point. This means black-hole masses less than $M_c$ will decay to zero but masses above $M_c$ will cascade. This can be seen from the solutions to the above
\begin{align}
    &\frac{u}{4}+M_i-M_c\coth^{-1}\left(\frac{M_i}{M_c}\right)= M(u)-M_c\coth^{-1}\left(\frac{M(u)}{M_c}\right) \quad  M_i > M_c,\\
    &\frac{u}{4}+M_i-M_c\tanh^{-1}\left(\frac{M_i}{M_c}\right)= M(u)-M_c\tanh^{-1}\left(\frac{M(u)}{M_c}\right) \quad  M_i < M_c,\\
    &M_c = 4\sqrt{\frac{|\xi|}{3}},
\end{align}

\subsection{Without azimuthal symmetry}
\begin{align}
    S_{in} &= 2\sum_{l}\int du  ~|a_{l, m}(u)|^2(2l+1)\left[2l\dot{M} + \frac{3\xi}{2l-3}M^{-2}-\frac{l}{2}\right],\\
    S_{out} &= 2\sum_{l}\int du  ~|b_{l,m}(u)|^2(2l+1)\left[2(l+1)\dot{M} + \frac{3\xi}{2l+5}M^{-2}-\frac{l+1}{2}\right].
\end{align}
This time in the path integral, we have
\begin{align}
    Z[M] &= \int [\mathcal{D}a_{l,m}][\mathcal{D}b_{l,m}] e^{iS} \notag\\
    &=\prod_{u\in\mathbb{R}}\frac{[\xi M^{-2}(4\dot{M}-1)^{2/3}(2\dot{M}+\frac{3\xi}{5}M^{-2}-\frac{1}{2})]^{1/2}}{\cos[2\pi g(\dot{M}, M)]}\frac{e^{4\left[\zeta(1)-1-\gamma\right](w^2_{+}+w^{2}_{-}-1/2)-2+8\zeta'(-1)}}{G(1+w_{+})^2G(2-w_{+})^2G(1+w_{-})^2G(2-w_{-})^2}.
\end{align}
Just like before there is a UV divergence in the effective action. Again treating this as our effective action
\begin{align}
    \Gamma_{div}\propto \int du ~g(\dot{M}, M)^2
\end{align}
we get the following equation of motion
\begin{align}
4\frac{d}{du}\left(\frac{M^{-1}}{4\dot{M}-1}\right)-M^{-2} = 0.
\end{align}
The above can be simplified to
\begin{align}
    &
    \left(\frac{4\dot{M}}{4\dot{M}-1}\right)^2 = 1+\alpha M^2\notag\\
    &\implies \dot{M} = \frac{1}{4\alpha M^2}(1+\alpha M^2)^{1/2}[(1+\alpha M^2)^{1/2}\pm 1].
\end{align}
If we look at
\begin{align}
    \dot{M} = \frac{1}{4\alpha M^2}(1+\alpha M^2)^{1/2}[(1+\alpha M^2)^{1/2}+ 1].
\end{align}
For $\alpha < 0$, the above does show thermodynamic behavior at $O(\alpha^{-1})$. The above has a stable critical point given by
\begin{align}
    M_r = \frac{1}{\sqrt{\alpha}}\quad
    \frac{d\dot{M}}{dM}\bigg|_{M=M_r}  < 0.
\end{align}
It means the above solution has a remnant of mass $M = M_r$. For the other sign configurations, the same critical point is unstable and therefore, has a cascading solution for masses $M > M_r$ and a decaying one for masses $M < M_r$ . So this has both non-thermodynamic evaporating and cascading black-hole configurations.\\

In summary, effective actions derived from path integral methods have both cascading or, in other words, accreting and evaporating configurations. The path integral gives rise to both expected and nontrivial accreting solutions. So far, we have only seen non-thermodynamic evaporating configurations. In the next section, we will see how the path integral also incorporates the expected thermodynamic evaporation. 
\subsection{Thermodynamic evaporation\\from s-wave dominance}
Black-holes are not perfect black-bodies due to the greybody factors $|B_l(p)|^2$ that appear in the mass loss rate formula which explicitly reads \cite{dewitt1975quantum}
\begin{align}
\frac{\mathrm{d} M}{\mathrm{~d} t}=-\frac{1}{2 \pi} \sum_{l=0}^{\infty} \int_0^{\infty}(2 l+1)\left|B_l(p)\right|^2 \frac{p}{\mathrm{e}^{8\pi M p }-1} \mathrm{~d} p
\end{align}
Explicit computations of the greybody factors show that the dominant contribution to the above comes from $l = s$ where $s$ is the spin of the field. For scalar fields, that is $l = 0$ which reads \cite{PhysRevD.13.198}
\begin{align}
    |B_0(p)|^2 = \frac{A}{\pi}p^2
\end{align}
where $A$ is the area of the black-hole event horizon. Hence, the mass loss rate formula explicitly becomes
\begin{align}
\frac{\mathrm{d} M}{\mathrm{~d} t}\approx-\frac{A}{2 \pi^2} \int_0^{\infty} \frac{p^3}{\mathrm{e}^{8\pi M p }-1} \mathrm{~d} p \propto -\frac{1}{M^2} \label{theva}
\end{align}
This precisely matches the Stephan-Boltzmann's law for a blackbody with temperature $T\propto 1/M$ and area $A \propto M^2$. This is referred to as the s-wave dominance. Hence, to derive the above dominant contribution from the path integral perspective, we will also assume s-wave dominance at late times both in the exterior and interior of the black-hole\cite{Hawking:1975vcx,Verlinde:1994dz,Brown:2024ajk}. Additionally, we also assume the interior of the black-hole to be non-dynamical. At the level of the path integral, this means that we have to work with 
\begin{align}
    &\varphi_{in}(u, r, \theta,\phi) = A~\psi^{in}_{0,0}(u, r,\theta,\phi),\notag\\
    &\varphi_{out}(u, r, \theta, \phi)=b_{0,0}(u)\psi^{out}_{0,0}(u, r,\theta,\phi).
\end{align}
Using this in the Scalar-Gauss-Bonnet action in Eq. (\ref{SGBA}) leads to 
\begin{align}
    S &= -A^2\xi\int du ~M^{-2} + \int du  ~b_{0,0}(u)^2\left[2\dot{M} + \frac{3\xi}{5}M^{-2}-\frac{1}{2}\right].
\end{align}
Then the path integral becomes
\begin{align}
    Z[M] &= \int [\mathcal{D}b_{0,0}]e^{iS} = e^{iA^2\xi\int du ~M^{-2}}\prod_{u\in \mathbb{R}} \left[2\dot{M} + \frac{3\xi}{5}M^{-2}-\frac{1}{2}\right]^{-1/2} \notag\\
    &=\exp\left[-iA^2\xi\int du ~M^{-2}+\frac{i}{2}\int du \ln\bigg|2\dot{M} + \frac{3\xi}{5}M^{-2}-\frac{1}{2}\bigg| \right].
\end{align}
If we look at the equation of motion of the above effective action, we get
\begin{align}
    2\ddot{M}-\frac{6\xi}{5}\dot{M}M^{-3} =\xi M^{-3}\bigg|2\dot{M} + \frac{3\xi}{5}M^{-2}-\frac{1}{2}\bigg|\left[2A^2\bigg|2\dot{M} + \frac{3\xi}{5}M^{-2}-\frac{1}{2}\bigg| -\frac{3}{5}\right].
\end{align}
One can see that the above can be solved by
\begin{align}
    \dot{M} = -\frac{3\xi}{10}M^{-2}, \quad A^2 = \frac{3}{5},
\end{align}
when $\xi > 0$ this is precisely the thermodynamically expected mass loss rate of a black-hole in Eq. (\ref{theva}).  This also confirms our path integral computations to be correct as well. However, notice that
\begin{align}
   & 2|\varphi^{in}_{0, 0}|^2 = |A|^2 T_H \quad 2|\varphi^{out}_{0, 0}|^2 = b_{0,0}^2(u)T_H\left(\frac{r_s}{r}\right)^2, \notag\\& 2|\psi^{in}_{0, 0}|^2\big|_{r=r_s} = 2|\psi^{out}_{0, 0}|^2\big|_{r=r_s} = T_H,
\end{align}
also has thermodynamic interpretation.  The dominant s-wave sector of the scalar fields represents the temperature profile of the spacetime due to the black-hole behaving like a heat source of temperature $T_H$. 

\section{Summary and Discussion}

We computed the effective theory for the mass parameter which allowed us to look at various black-hole configurations just by computing their equations of motion. We showed that the equation of motion also has cascading or accreting solutions and one of them is consistent with a well-known accretion formula \cite{bondi1952spherically} in the large mass regimes. Non-thermodynamic evaporating solutions with highly-stable remnant endstages emerge depending on the sign of the parameters of the scalar theory. Thermodynamic decay only occurs when the matter couples non-minimally through the SGB term with the assumption of s-wave dominance. This demonstrates that black-hole evaporation or cascade is dependent on the matter action as was precisely hinted in $2D$ computations of \cite{Leahy:1983vb}. The requirement of the SGB term in order to reproduce thermodynamic behaviour implies an underlying topology change as a possible cause for the thermodynamics. This is something we leave for our future investigations. In this computation, we also established the scope of the thermodynamic regime of black-holes to SGB interactions as far as bosonic matter is concerned. It is possible that there is a class of matter interactions where the thermodynamic regime is more dominant and fundamental. If true, one can postulate a thermodynamic constraining principle for the kind of matter interactions that can be allowed in gravity. That would require repeating this computation for a wide variety of matter configurations which we intend to undertake in our future studies.\\



\section*{Acknowledgements}
AM would like to acknowledge the moral support of Hare Krishna Movement, Pune, India. AM was supported by the NRF grant funded by the Korea government (MSIT) (No. 2022R1A2C1003182). This research is dedicated to the people of Republic of India, Republic of Colombia, Macao Special Administrative Region of the People's Republic of China and the Republic of Korea for their steady support of research in theoretical science. AM would like to thank Dr. Vinay Malvimat (Kyunghee University) for various fruitful discussions on black-hole information paradox and thermodynamics. IA and AM both would like to thank the organizers of workshop titled ``Quantum Gravity and Information in Expanding Universe" held at YITP, Kyoto in February, 2025 where this collaboration began.\\\\
\textbf{Conflict of interest:} The authors report no conflict of interest.\\
\textbf{Data availability:} No data was used or generated in this research.

\appendix
\section{Spherical mode functions in a time-dependent, one-way permeable boundary}\label{movBound}
Consider the following PDE 
\begin{align}
    r^2\vec{\nabla}^2\psi_{l,m} = -k^2r^2\psi_{l,m}
\end{align}
The above is solved by
\begin{align}
    \psi_{l,m}(r,\theta,\phi) = f(kr)Y_{l, m}(\theta,\phi)
\end{align}
such that
\begin{align}
    \frac{d}{dr}\left(r^2\frac{d f}{dr}\right)+(k^2r^2-l(l+1))f = 0
\end{align}
The above ODE has two independent solutions given by
\begin{align}
    f(r) = A~j_l(kr) + B~y_l(kr)
\end{align}
which are spherical Bessel functions. Given a spherical, one-way permeable boundary at $r = 2M$, the solution must take the following form
\begin{align}
f(r) = f_{out}(r) \Theta(r - 2M) + f_{in}(r) \Theta(2M-r)\label{ansatz2}
\end{align}
where in the above the time-dependence $u$ is made implicit. Permeability means that the first derivative of the field must not have divergences i.e.
\begin{align}
    f'(r) = f'_{out}(r) \Theta(r - 2M) + f'_{in}(r) \Theta(2M-r) + [f_{out}(2M)-f_{in}(2M)]\delta(r-2M)
\end{align}
the term proportional to the $\delta$-function must vanish. Hence, we must have
\begin{align}
    f_{in}(2M) = f_{out}(2M)
\end{align}
which leads to
\begin{align}
    f_{in}(r) = \frac{1}{\sqrt{N}}~y_l(2Mk)j_l(kr) \quad f_{out}(r) = \frac{1}{\sqrt{N}}~j_l(2Mk)~y_l(kr) \label{sphmodfn3}
\end{align}
The above choice avoids divergences at $r=0$ and $r \to \infty$. 
Now, we normalize the field as follows
\begin{align}
    \int_0^{\infty} dr f(r)^2 = \int_{0}^{2M}f_{in}(r)^2 +\int_{2M}^{\infty} dr ~f_{out}(r)^2   = 1
\end{align}
where in the above we have made use of\cite{Poisson:2011nh}
\begin{align}
    \Theta(x)^2 = \Theta(x) \quad \Theta(x)\Theta(-x) = 0
\end{align}
This leads to
\begin{align}
    N = y_l(2Mk)^2\int^{2M}_0 dr ~j_l(kr)^2 + j_l(2Mk)^2\int^{\infty}_{2M} dr ~y_l(kr)^2 
\end{align}
which determines $N$ as a function of $k, l, M$. Now, we want $f(r)$ to represent bound states. The reason why we want this will be made clear below. Since, bound states by definition are time-independent states\cite{ruelle1969remark}, hence, we must impose $\partial_{u}f(r) = 0$ which leads to
\begin{align}
    \partial_uf(r) &= \partial_u f_{out}(r) \Theta(r - 2M) + \partial_u f_{in}(r) \Theta(2M-r) + 2\dot{M}[f_{out}(2M)-f_{in}(2M)]\delta(r-2M)\notag\\
    &= -\frac{\dot{N}}{2N}f(r)+ \frac{2k\dot{M}}{\sqrt{N}}\left[y'_l(2Mk)j_l(kr)\Theta(r - 2M)+j'_l(2Mk)y_l(2Mk)\Theta(2M-r)\right]\notag\\
    &= 2k\dot{M}\left[-\frac{f(r)}{N}\left(y'_{l}(2Mk)y_{l}(2Mk)\int^{2M}_0 dr ~j_l(kr)^2+j'_{l}(2Mk)j_l(2Mk)\int^{\infty}_{2M} dr ~y_l(kr)^2\right)+\right.\notag\\&~~~~~\left.\frac{1}{\sqrt{N}}\bigg(y'_l(2Mk)j_l(kr)\Theta(r - 2M)+j'_l(2Mk)y_l(2Mk)\Theta(2M-r)\bigg) \right] = 0
\end{align}
Notice how either $\dot{M} = 0$ or $k = 0$ both can fulfill the conditions for $f(r)$ to be a bound state. But $\dot{M} = 0$ precisely reduces to the well-known case of Hawking's computation where the black-hole mass $M$ is indeed taken to be classically constant in time \cite{Hawking:1975vcx}. In other words, Hawking's computation implicitly works with bound states by working with a black-hole metric of fixed mass $M$. In fact, the radial part of the mode expansion used in the Hawking's computation can precisely be recast into Eq. (\ref{ansatz2}). As black-hole evaporation require such bound states, we impose this condition as well. Since, for our case $\dot{M} \neq 0$ because of the Vaidya metric, we simply take $k = 0$ in Eq. (\ref{sphmodfn3})to implement the bound state criterion which precisely lead to Eq. \ref{sphmodefn}. This can be physically interpreted as infinite redshift at the event horizon. We will not discuss the term in the square brackets being zero which imposes a nontrivial constraint relation on $k, l, M$. Whether or not this leads to anything physically significant or interesting is something we relegate to our future studies.
\bibliographystyle{unsrt}
\bibliography{QGBib}

\end{document}